\documentclass[article]{IEEEtran}
\usepackage{geometry}
\usepackage{amsmath, amsfonts, amssymb}
\usepackage{stfloats}
\usepackage{graphicx}
\usepackage{amsthm}
\theoremstyle{definition}
\newtheorem{definition}{Definition}
\usepackage{mathtools}
\usepackage{algorithm}
\usepackage{algpseudocode}
\usepackage{float}
\usepackage{soul}
\usepackage{endnotes}
\usepackage[font={footnotesize}]{caption}
\usepackage{subcaption}
\usepackage{booktabs}
\usepackage{adjustbox}
\usepackage{pdflscape}
\usepackage{makecell}
\usepackage{todonotes}
\usepackage{hyperref}
\usepackage{cleveref}
\usepackage{xcolor}
\usepackage{float}
\algrenewcommand\algorithmicreturn{\textbf{return}}

\title{Unpublished Draft: A Post-Processing Approach to Fairness in Tie-Aware Rankings}
\author{ \and  \and }
\author{\IEEEauthorblockN{Somya Nigam,
Johan Springael,
Kenneth Sörensen}\\

\IEEEauthorblockA{Department of Engineering Management, ANT/OR Research Group,
University of Antwerp, Belgium.\\
somya.nigam@uantwerpen.be,
johan.springael@uantwerpen.be,
kenneth.sorensen@uantwerpen.be,
}}

\date{} %
\newcommand{\smallAccuracy}{~84.4-86.6}

\newcommand{\mediumAccuracy}{~26.6-28.8}

\begin{document}
\raggedbottom
 
\maketitle
\title

\sethlcolor{yellow} 
\begin{abstract}
 The problem of finding a fair consensus ranking is an active research topic in the domain of fair rank aggregation and has been well studied; however, existing studies predominantly consider both the input rankings and the output ranking to be permutations where elements are always strictly ordered. In practice, however, a ranking with ties is far more common. This study bridges this gap by presenting a post-processing approach to determine the closest fair consensus ranking when an unfair tie-aware consensus ranking is provided. It proposes an exact algorithm and a fast heuristic to achieve this.
\end{abstract}


\section{\textbf{Introduction}}
Rankings are a pervasive mathematical tool used to define preferences and priorities among a set of items, places or candidates. Across various domains, decision-making heavily relies on preferences elicited from subject-matter experts. These preferences are often expressed in the form of rankings. These rankings are then collectively analysed to produce a single consensus ranking, which is essentially a classical rank aggregation problem that has been studied for over three centuries. This problem was first studied in the context of elections \cite{young1988condorcet}, later giving rise to the field of economics known as social choice theory \cite{8defed17-142f-3287-bede-acf83813936a}. Its scope and applications, however, are not limited to elections; it has found considerable traction in information retrieval \cite{dwork2001rank},\cite{farah_outranking_2007}, sports competitions \cite{davenport_ranking_nodate},\cite{truchon_extension_nodate}, collaborative filtering \cite{tang_bordarank_2016}, bioinformatics \cite{patel_hierarchical_2013}, and other domains. The complexity of the problem rises even when the number of base rankers is greater than or equal to 4 \cite{dwork2001rank}, thereby making it an NP-hard problem. Thus, numerous methods have been proposed in the literature to compute a consensus ranking in polynomial time.

Decisions based on rankings can have a significant impact on human lives, particularly in the face of resource scarcity. This scarcity is not evenly distributed even across the globe \cite{chancel2022world}.
In modern societies across countries, income inequality persists and continues to rise despite significant economic and technological advancements. This phenomenon is largely driven by political and institutional choices that govern how societies are organised \cite{chancel2022world}. Marginalisation and oppression have taken many forms historically, and their effects remain evident today. For instance, the 2022 report \cite{chancel2022world} shows that, across all countries, women receive only one-third of total labour income and remain severely underrepresented in high-paying jobs and leadership positions.

Racial discrimination and violence have occurred in various historical and national contexts; for example, in the United States, African Americans experienced some of the most severe forms of discrimination and violent oppression in the period following the Second World War \cite{JimCrowMuseumWhatWas, initiative2020lynching}. Owing to these historical and structural power imbalances, particularly along gender and racial lines, affected communities have disproportionately remained in lower-income and wealth brackets \cite{PewWealthGaps2023, chancel2022world}.
When rankings are constructed over individuals, communities, or countries, decisions based on traditional rank aggregation methods often overlook or inadvertently reinforce such underlying disparities. To counteract these effects, researchers have begun incorporating \textit{fairness} considerations into ranking methodologies \cite{yang_measuring_2016, singh_fairness_2018, kuhlman_rank_2020, zehlike_fairness_2023-1}.

The definition of fairness, however, is not straightforward and is inherently task-specific \cite{friedler_impossibility_2016}. For an illustrative perspective, one may turn to Rawls’ theory of justice, which advocates designing society so that the least advantaged members of society are better off, independent of individual self-interest \cite{freeman_original_2023}. In real-world settings, many institutions have adopted corrective mechanisms aimed at compensating for historical disadvantage, such as gender quotas in high-paying positions \cite{europarl2022corporateboards} or targeted investments intended to promote the upward mobility of racial groups previously affected by systemic oppression \cite{Biden2025Record}.

Moreover, the adoption of data-driven algorithms by large corporations in today's time could potentially amplify existing societal discrimination through the use of underlying biased data.\cite{barocas2016big}.
Recent years have seen a widespread deployment of AI algorithms across a variety of domains \cite{pena_human-centric_2023}. Several well-documented cases (see \cite{JeffLarson}, \cite{Dastin2018}) demonstrate that the use of machine learning models by large corporations has had adverse effects on various demographic groups \cite{pena_human-centric_2023}; these instances have led to a substantial body of research on fairness-aware machine learning \cite{chinta_ai-driven_2025, rabonato_systematic_2025, zhang_fairness_2023}.

In the context of ranking systems, LinkedIn® has used a fairness-aware re-ranking framework in its  LinkedIn Recruiter™ product. When recruiters or hiring managers search for talent to meet their specific requirements, the system first retrieves candidates who satisfy the criteria and scores them using the machine learning models. The scored list then gets post-processed by a re-ranking algorithm, which then provides a gender-representative ranking to the recruiter,  \cite{geyik_fairness-aware_2019}; this is done to prevent any existing over/under-representation of any gender group.

More recently, a new branch of research focusing on fairness in the rank aggregation task (FRA) [\cite{wei_rank_2022},\cite{chakraborty_improved_2025}, \cite{kuhlman_rank_2020}, \cite{cachel_mani-rank_2022}, \cite{cachel_prefair_2024}, \cite{chakraborty_fair_2022},\cite{chakraborty2023fairrankaggregation}, \cite{cachel_fairer_2023}] has emerged. The primary motivation for incorporating fairness in this task is to mitigate the risk of potential biases that may arise in the consensus ranking and to ensure that no particular group is deprived of any opportunities due to their misrepresentation in the consensus ranking. 
To the best of our knowledge, existing studies have primarily focused on permutations (i.e. strict or complete rankings), both as input rankings and as the consensus ranking. Many of these approaches are centred around finding the closest fair ranking once a ranking is already provided, which is indeed an interesting problem in its own right.
In practice, however, ranking with ties (henceforth tie-aware ranking, used interchangeably) frequently occurs, because it is not always possible for decision-makers to strictly order the set of items \cite{kendall1945treatment}. And even if we break the ties arbitrarily, what could happen is that sometimes one of the groups could get an added advantage over the other, which is equally unfair.
Therefore, this study seeks to answer a key research question: Given a consensus ranking with ties, how can we determine the closest fair ranking under a specific fairness notion?



In this paper, we make the following contributions:
\begin{itemize}
    \item We provide a first exact algorithm which accepts a tie-aware ranking or even permutation as input and optimally determines the closest fair ranking under the pairwise statistical parity notion of fairness.
    \item Our exact algorithm can accommodate a ranking(s) of size $\ge$ 100 and provides an optimal solution for any fairness threshold. It is also equipped to handle multiple rankings simultaneously. 
    \item We also present a heuristic to find a reasonable closest fair ranking.
\end{itemize}
The rest of the paper is organised as follows. In \cref{sec:related}, we provide an overview of the relevant literature. This is followed by \cref{sec:preliminaries}, where we introduce the notations and preliminaries used to describe the problem.
We then formally define the problem (\cref{sec:problem}), describe the data generation process (\cref{sec:data}), and present our methodology (\cref{sec:method}) alongside experimental results (\cref{sec:experiments}). 
Finally, we discuss the limitations of our approach and conclude the paper in \cref{sec:conclusion}.
\section{Related Work}
\label{sec:related}

This section provides an overview of the topics covered in our comprehensive literature survey, which together provide the foundation for understanding the problem at hand. The section first discusses Rank Aggregation with or without ties,
followed by Fair Ranking, which also covers Fair Rank Aggregation.

\subsection*{\textbf{Rank Aggregation}}\vspace{-0.5em}
\subsection*{\textit{Ranking without Ties (a.k.a. complete/strict rankings or permutations)}}

As stated earlier in the introduction, rank aggregation is a complex problem; therefore, numerous methods have been devised to address this task. Kemeny optimal aggregation (based on Kendall’s distance metric) \cite{kemeny1959mathematics} is almost synonymous with Rank Aggregation, as it is a commonly used objective function because it satisfies some important properties, such as neutrality, consistency, and the Condorcet criterion \cite{young1978consistent}. However, it violates Arrow’s independence of irrelevant alternatives axiom \cite{maskin_arrow_2014},\cite{kemeny1959mathematics} and is NP-hard, even though the number of base rankers is just 4 \cite{dwork2001rank}. Other distance metrics exist, such as Spearman’s Footrule \cite{diaconis1977spearman}, which is at least the Kendall’s Tau Distance and could be at most twice the Kendall’s Tau Distance (see \cite{diaconis1977spearman} for the precise relationship). 

Several other approximate algorithms, such as Borda’s method \cite{borda1781m}, Pick-a-Perm, KWIKSORT (formulated as a weighted feedback arc set problem, \cite{coleman_ranking_2009,ailon_aggregating_2008},  Markov chain-based methods (MC1, MC2, MC3 and MC4) \cite{dwork2001rank}, Copeland’s method \cite{copeland1951reasonable,schalekamp2009rank}, and many more. Exact optimisation formulations are given in \cite{conitzer2006improved}, \cite{davenport2004computational}. 

\subsection*{\textit{Ranking with Ties}}
Rank Aggregation, considering ranking with ties, is also NP-hard \cite{cohen2011using}. Several studies in the literature have addressed the problem of aggregating and comparing partial rankings.
Fagin et al. were among the first to tackle this problem systematically. They presented four metrics to compare partial rankings, obtained by generalising the well-known Kendall’s tau and Spearman’s footrule distances, originally defined for complete permutations. To account for ties, they introduced a penalty p to the originally defined Kendall Tau Distance, which applies a penalty parameter $p$ to any pair of elements that are tied in one ranking but strictly ordered in the other \cite{fagin_comparing_2004}. In their subsequent work \cite{fagin2006comparing}, they formally showed that when 0<$p$<1/2, the generalised distance metric is a near metric; it is a true metric when $1/2\le p \le1$ and not a metric when $p$=0.
They showed that the four metrics are equivalent, meaning they differ only by a constant multiple. The formal definition of the generalised Kendall Tau distance is in the ~\ref{sec:preliminaries} section.
Ailon \cite{ailon_aggregation_2010}  introduced some approximation algorithms called ‘RepeatChoice’ and  ‘LPKwikSorth’. Both algorithms produce a permutation even when they accept partial rankings. ‘RepeatChoice’ is a randomised expected 2-factor approximation algorithm and ensures ties are broken progressively using a * refinement operator, while ‘LPKwikSorth’ is a 3/2-factor approximation algorithm.
In some cases, it may be desirable to obtain a partial ranking as the aggregated result. Cohen-Boulakia et al. studied this specific problem in the context of biological data and proposed a heuristic called ‘BioConsert’. This heuristic makes use of two specific operations: ‘changeBucket’, which allows moving an item into a different existing bucket, and ‘addBucket’, which enables placing an item into a new bucket altogether \cite{cohen2011using}.
Brancotte et al. developed an exact algorithm for finding an optimal consensus ranking in the presence of ties, using Integer Linear Programming (ILP) \cite{brancotte_rank_2015}. We adopt this in our approach; further details are provided in section~\ref{sec:method}.
\subsection*{\textit{Fair Ranking}}
Rankings often suffer from position bias or exposure; for instance, in hiring, recruiters tend to pay more attention to candidates ranked at the top, with attention decreasing for those ranked lower. Hence, researchers \cite{singh2018fairness} provided a framework to express fairness constraints in terms of exposure allocation with an overall aim to maximise utility to the user. 
They considered constraints of the form: Demographic parity, Disparate Treatment and Disparate Impact and provided a Linear program and BvN decomposition to obtain probabilistic rankings which are fair in expectation.
Several post-processing algorithms for rankings have been proposed in the literature. For example, Celis et al. \cite{celis_ranking_2018} studied the constrained ranking maximisation problem, where the idea is 
to place \textit{m} items into \textit{n} positions such that the profit associated by placing those \textit{m} items into specific \textit{j} positions is maximised. Along with this, the items belong to some sensitive groups; to ensure that none of the sensitive group dominates the ranking, an additional constraint was considered that forces the items to adhere to the lower-bound and upper-bound thresholds in the top-\textit{k} positions of the ranking. 
The algorithm FA*IR presented in \cite{zehlike_fair_2017} aims to select the \textit{k} best items from the pool of \textit{n} candidates, such that all the qualified candidates are always ranked higher than the unqualified candidates and satisfying the group fairness criterion. The group fairness criterion requires that the proportion of the protected candidates in every prefix of the top-k ranking is statistically no smaller than a given minimum threshold.
Other algorithms proposed in \cite{geyik_fairness-aware_2019} aim to solve a similar problem and provide an extensive framework.
We have highlighted a few representative studies in this section. For a more comprehensive and detailed overview of the fairness in ranking literature, we refer the reader to the survey \cite{zehlike_fairness_2023,pitoura_fairness_2021}.
Kuhlman et al. \cite{kuhlman_rank_2020} were the first to study fairness in rank aggregation. They focus on a single protected attribute with binary groups/categories. Their exact algorithm, Fair-ILP, finds a consensus ranking under the pairwise statistical parity criterion but can only handle permutations and is limited to fewer than 100 items. They also proposed Fair-BB, a Branch-and-Bound method for fair rank aggregation that uses a lower bound on the distance from the base rankings to any fair consensus ranking so as to avoid costly backtracking. Their approximate algorithm, Fair-Post, takes any unfair ranking as input and then corrects it for fairness. 

Other studies, such as \cite{wei_rank_2022},\cite{chakraborty_fair_2022},\cite{chakraborty_improved_2025},\cite{chakraborty2023fairrankaggregation} study a similar FRA problem but under the proportionate fairness notion with the protected attribute having multiple groups.

To our knowledge, the studies listed above, as well as all prior work on fairness in individual rankings and rank aggregation, are restricted to permutations as input rankings. In contrast, we consider a ranking with ties as an input to our algorithms.

Cachel et al. \cite{10.1145/3715275.3732042} also identified a similar knowledge gap; however, their approach differs significantly from ours. Their study focuses on an in-process approach to fairness; for that matter, they consider ratings, which are a special case of partial rankings \cite{ailon_aggregation_2010}, and the output of their fair-rated aggregation problem is always a permutation. In our work, no such condition is imposed on the output ranking and thus, tie-breaking is applied where appropriate.

Another line of work \cite{aledo2025consensus} highlights a similar gap; however, their study is situated in a different field and employs a distinct mathematical formalism. Moreover, their approach does not explicitly enforce a fairness measure; instead achieving fairness through a set of weighted rankings as output. Due to these structural differences, a direct comparison with our results would be mathematically challenging.

\section{\textbf{Preliminaries and Formalism}}
\label{sec:preliminaries}
\subsection*{Ranking Representations}
Let there be an alternative set $A$ of items(candidates, countries, etc.) such that $|A| = n$, where $n \in \mathbb{N}$. 
Then, for any $a_i \in A$, a strict ranking (a.k.a. permutation) can be represented as 
$\pi = (a_1 \prec a_2 \prec \dots \prec a_n)$, 
implying that $a_1$ is placed at a more advantageous position than $a_2$.

However, a tie-aware ranking  
can be formalized as a \emph{bucket order}. Formally, a bucket order is a transitive binary relation $\triangleleft$ over a set partitioned into buckets $B_1, B_2, \dots, B_t$. For any items $a_1$ and $a_2$, we have $a_1 \triangleleft a_2$ if and only if $a_1 \in B_i$, $a_2 \in B_j$, and $i < j$, indicating that $a_1$ precedes $a_2$ in the given ranking. 
The members of a given bucket are considered ``tied", 
As a special case, a strict ranking is also a bucket order, where each bucket is of size 1 \cite{fagin_comparing_2004}.
In other words, 
a bucket order is formed by grouping the items of a ranking and sorted according to their rank values. If a given rank value is assigned to only one item, the corresponding bucket contains that single item and has size $1$. If a rank value is shared by $k$ items, then the corresponding bucket contains those $k$ elements and has size $k$.The total number of buckets in the bucket order will be equal to the maximum value of the original ranking.

\textit{Note that a tie-aware ranking can also be called a partial ranking \cite{brancotte_rank_2015}. Partial ranking sometimes includes incomplete rankings or ratings as well \cite{ailon_aggregation_2010}. In this article, any occurrence of \textit{partial ranking} simply refers to a ranking with ties. }
\subsection*{Distance Metrics}
There are various distance metrics to capture the disagreement between any two rankings. The most popular choices when comparing two permutations are Kendall-Tau Distance \cite{kendall1938new} and Spearman's Footrule Distance \cite{diaconis1977spearman}. \cite{diaconis1977spearman} proved an equivalence relation between these two metrics.

To compare rankings with ties, Fagin \emph{et al.}\cite{fagin_comparing_2004} introduced a generalised Kendall-Tau distance metric, with penalty $p$ to account for ties. As pointed out earlier, when $1/2\le p\le1$, it is a true metric. Following Brancotte \emph{et al.}\cite{brancotte_rank_2015}, we set $p=1$ as well.

\begin{definition}
Let $\sigma_1$ and $\sigma_2$ be two tie-aware rankings.  
The generalized Kendall–Tau distance $G(\sigma_1, \sigma_2)$ is defined as:
\begin{align}
G(\sigma_1, \sigma_2) 
&= \sum_{i=1}^{n} \sum_{j=i+1}^{n} 
\mathbf{1}_{e_{ij}},
\\[6pt]
\text{where,} \quad 
e_{ij} &= 
\Big[\,
(\sigma_1(i) < \sigma_1(j) \land \sigma_2(i) > \sigma_2(j)) \notag\\
&\qquad\lor\, (\sigma_1(i) > \sigma_1(j) \land \sigma_2(i) < \sigma_2(j)) \notag\\
&\qquad\lor\, (\sigma_1(i) = \sigma_1(j) \land \sigma_2(i) \neq \sigma_2(j)) \notag\\
&\qquad\lor\, (\sigma_1(i) \neq \sigma_1(j) \land \sigma_2(i) = \sigma_2(j))
\,\Big]. \notag
\end{align}

Here, $\sigma_k(i)$ and  $\sigma_k(j)$ denote the \emph{rank value} assigned to items~$i$ and ~$j$ in ranking~$\sigma_k$, for $k \in \{1,2\}$.

In other words, the generalized Kendall–Tau distance quantifies the pairwise discrepancies between two tie-aware rankings. Two types of disagreement can arise, as described below.

Scenario 1: If two items \(i\) and \(j\) are strictly ordered in one ranking, with \(i\) preceding \(j\), and this order is reversed in the other ranking (i.e., \(j\) precedes \(i\)) (or vice-versa), then the distance metric detects this inversion and assigns a cost of 1.

Scenario 2: If the items \(i\) and \(j\) are tied in one ranking but strictly ordered in the other ranking (or vice versa), the distance metric also identifies this as a disagreement and assigns a cost of 1.
\end{definition}

\subsection*{Fair Ranking}
In our problem setting, we are provided with sensitive information about the items in our ranking.
This sensitive information is often referred to as a \textit{protected attribute} such as race, gender, religion, etc., which is deemed irrelevant to the task at hand and cannot be used as a basis to make decisions about the items under study. Items with the same value of the protected attribute are said to be in a \textit{group}.

Formally, we focus on \textit{group fairness} (see \cite{dwork_fairness_2011} for formal definition), where each element of $A$ is associated with a \textit{protected attribute}. $G(i)$ that partitions $A$ into $g$ disjoint groups. If $g=2$, then we have a binary protected attribute, with one group representing the protected (or minority) group and the other representing the majority group.  
If $ g> 2$, then we have multiple groups in the protected attribute. 
In this study, we consider a binary protected attribute, namely $G_1$ and $G_2$, the assignment of which group is protected and which is non-protected depends entirely on the given unfair ranking.
    

There is a plethora of fairness measures 
defined in the literature; however, there is no clear criterion to help decide how a measure should be chosen. Schumacher et al. \cite{schumacher_properties_2025} argued that the choice of measure depends on the notion of fairness (which isn’t universal \cite{friedler_impossibility_2016}) to be met in a given application.
The notion of fairness often draws from philosophical theories such as Equality of Opportunity (EO) and Substantive EO \cite{schumacher_properties_2025},\cite{zehlike_fairness_2023}. The former emphasises the importance of personal (or native) qualifications and seeks to minimise the impact of circumstances and arbitrary factors (such as race or family background) on individual outcomes. For instance, in the case of hiring, candidates with the required qualifications and experience are to be chosen for the job without regard to attributes such as race or gender \cite{zehlike_fairness_2023}. The latter acknowledges that the necessary qualifications may themselves result from arbitrary factors (such as socio-economic background) and thus allows for fair access to attain the required qualifications \cite{zehlike_fairness_2023},\cite{heidari_moral_2019}. In other words, it aims to make the future prospects of people comparable  \cite{zehlike_fairness_2023}. Researchers adopt several such worldviews (see \cite{friedler_impossibility_2016}).
To begin with, we consider the \textit{pairwise statistical parity measure} of fairness, which aims to equalize outcomes across protected and non-protected groups \cite{narasimhan_pairwise_2020}, \cite{dwork_fairness_2011}. In other words, in a ranking problem, candidates from the protected group or the non-protected group have an equal chance of being ranked in top positions. For our implementation, we use the constraint formulation described in \cite{kuhlman_rank_2020}.




\section{\textbf{Problem Definition}}
\label{sec:problem}
The formal definition of finding the closest fair ranking is as follows.

\begin{definition}
 Given a tie-aware consensus ranking $\pi$ of $n \in \mathbb{N}$ items, the closest fair ranking $\pi^*$ minimises the distance $G(\pi,\pi^*)$ under the pairwise statistical parity notion.  
 \end{definition}

\textbf{Problem Complexity :} This problem is NP-Hard, and the total number of possible solutions is given by $\sum_{g=1}^{n} g! \, S(n,g)$, where g is the number of buckets,n is the number of items in the rankings, and S(n,g) denotes the Stirling number of the second kind. Unlike when dealing with permutations, where the problem is n!, our problem is comparatively harder.
\section{\textbf{Data Generator}}
\label{sec:data}
To carry out the study, we create a synthetic dataset. 
For this purpose, we design an algorithm that randomly generates a ranking(s) of a finite 
number of items ($n$) with a specified proportion of ties.
The items belong to two 
groups, \textit{$G_1$} and \textit{$G_2$}, and are randomly assigned to these groups. The ranking obtained is random in nature; if it happens to be statistically fair, it will occur purely by chance.

\begin{algorithm}[H]
\caption{Ranking Instances}
\label{alg:cap1}
\begin{algorithmic}[1]

\State \textbf{Input:} n, seed, rank\_number, tie\_prob
\State \textbf{Output:} ranking\_dataset
\State set\_seed(seed)
\State Initialize ranks as a $\mathit{rank\_number} \times n$ matrix filled with zeros
\State Initialize empty list total\_rank
\State Initialize empty dataset df
\State Initialize a list Groups with elements G1 and G2
\State additional\_size= n-len(Groups)
\State Generate a RandomSample of length additional\_size with replacement from Groups and store it in additional
\State Groups ← Concatenate(Groups, additional)
\State Create list Items with item labels $\{A_1, A_2, \ldots, A_n\}$ 
\State Assign Groups and Items as columns in df
\For{$k \gets 1$ \textbf{to} $rank\_number$}
    \State rank = ranks[k]
    \State shuffled\_idx = permutation(n) \Comment{{\footnotesize Generate a permutation of size n}}
    
    \State current\_rank = 1
    \State rank[shuffled\_idx[0]] = current\_rank \Comment{{\footnotesize Assign current\_rank at some position in rank}}

    \For{$i \gets 1$ \textbf{to} $n$}
        \If{random\_number $\le$ tie\_prob}
            \State rank[shuffled\_idx[i]] = current\_rank
        \Else
            \State current\_rank = current\_rank + 1
            \State rank[shuffled\_idx[i]] = current\_rank
        \EndIf
    \EndFor

    \State total\_rank.append(current\_rank)
\EndFor

\For{$i \gets 1$ \textbf{to} len(total\_rank})
    \State df["Ranks\_i"] = total\_rank[i]
\EndFor

\State \Return df

\end{algorithmic}
\end{algorithm}

\section{\textbf{Methodology}}
\label{sec:method}
We have framed our problem as a binary integer linear program (BILP). In particular, we build upon the formulation provided by Brancotte et al. \cite{brancotte_rank_2015} as a backbone, adapting it by adding a fairness constraint to satisfy our fairness requirements.
As discussed earlier, we consider the pairwise statistical parity notion of fairness, as described in \cite{kuhlman_rank_2020} and incorporate it as a constraint in our linear program. Our exact algorithm, EFTA-BILP, short for: \underline{E}xact \underline{F}air \underline{T}ie \underline{A}ware - \underline{B}inary \underline{I}nteger \underline{L}inear \underline{P}rogram, is formulated as follows.

%
\begin{equation}
\begin{aligned}
&\text{Min Z=} \sum_{a_1,a_2}w_{a_2\leq a_1}x_{a_1<a_2}+w_{a_1 \leq a_2}x_{a_2<a_1} +\\
& \quad(w_{a_1<a_2}+w_{a_2<a_1})x_{a_1=a_2}  
\end{aligned}
\end{equation}




\noindent
\text{Subject to}
\begin{align}
&x_{a_1<a_2} + x_{a_2<a_1}+ x_{a_1=a_2} =1 \\               
&x_{a_1<a_3} - x_{a_1<a_2}- x_{a_2<a_3} \ge-1 \\
&2x_{a_1<a_2}+ 2x_{a_2<a_1} + 2x_{a_2<a_3} + 2x_{a_3<a_2}- \nonumber \\
&\quad x_{a_1<a_3}-x_{a_3<a_1} \ge 0 \\
\newcommand{\Factor}{\left( \frac{1}{|G_1||G_2|} \right)}
&\lvert\sum_{\substack{a_1 \in G_1 \\ a_2 \in G_2}} (x_{a_1<a_2}-x_{a_2<a_1})\rvert\le \epsilon  (|G_1||G_2|)
\end{align}
where, $[x_{a_1<a_2}, x_{a_2<a_1}, x_{a_1=a_2}] = 0$ or $1$ and $\epsilon \in [0,1]$.

\!\!\textbf{Explanation:}The decision variables $x_{a_1<a_2}$, $x_{a_2<a_1}$ and $x_{a_1=a_2}$ represent the position of the items $a_1$ and $a_2$ in the resultant fair ranking. If $a_1$ is placed at a better position than $a_2$ then  $x_{a_1<a_2}$ is set to 1, otherwise $x_{a_2<a_1}$ is 1; else when both $a_1$ and $a_2$ are tied then $x_{a_1=a_2}$ is 1.\\
Equations (2), (3), (4) and (5) represent the ordering constraints, transitivity constraints, consensus ranking constraints with ties, and parity constraints, respectively.
The $w$ variables comprising the objective function count the number of times $a_1$ has been ranked before/after/equal to $a_2$ in a given ranking or a set of rankings; together with the $ x$ decision variables, the objective function aims to minimise the generalised Kendall-Tau distance.

\subsection*{Heuristic}
We also propose a fast heuristic to handle instances where the exact algorithm fails to provide a solution within a reasonable time. Our heuristic employs a hybrid metaheuristic framework that incorporates Sequential Local Search (SLS), Tabu Search, Simulated Annealing (SA), and the Epsilon-Greedy Policy. A high-level overview of the heuristic is as follows. 
\\We have designed our heuristic at the level of buckets (as explained in \cref{sec:preliminaries}) 
, thereby exploiting the ties information in the ranking. This gave us a greater advantage because, by working at the bucket level, we could confine the search space.


After carefully examining the exact algorithm solution, we consider three move types in our algorithm, namely \textbf{Swap}, \textbf{CreatingTies}, and 
\textbf{BreakingTies}. The definitions are as follows.
\begin{definition}
Let $\sigma$ be a ranking of $n$ items which can be expressed in the form of k  buckets, where \[
k =
\begin{cases}
< n & \text{if at least one pair of items is } tied, \\
n & \text{otherwise.}
\end{cases}
\]
And let there be two buckets $B_i$ and $B_j$ having rank values i and j respectively, where $i<j$, i, j = 1, 2, \ldots, k.\\
\textbf{Swap}: An operation which, when applied to any two buckets $B_i$ and $B_j$ with $i < j$, exchanges their associated rank values by assigning rank $j$ to all items in $B_i$ and rank $i$ to all items in $B_j$.\\
\textbf{CreatingTies}: An operation that, when performed on any two buckets $B_i$ and $B_j$ with $i < j$, combines $B_i$ and $B_j$ into a single bucket $B_i'$ and assigns the rank $i$ to $B_i'$.\\
\textbf{BreakingTies}: An operation that, when applied to a bucket $B_i$ containing multiple items, divides it into two or more disjoint buckets, say $B_k$ and $B_l$, such that the rank of $B_k$ is $i$, the rank of $B_l$ is $i+1$, and $B_i = B_k \cup B_l$.

\end{definition}
As follows from the definitions, designing the algorithm at the level of buckets also has another advantage: we can apply operations to multiple items at once (provided the buckets contain multiple items), which significantly reduces the overhead and is a relatively easier and more structured approach compared to working at the level of ranking.
Our algorithm runs for several iterations for \texttt{max\_iterations} and comprises the following modules.
\subsection*{SLS \& Tabu Search Phase }

The process begins by extracting the indices of the buckets and the items' information within them on which the operation \textbf{Swap},
is to be applied. The operation runs if the items are not in the \texttt{tabu\_list\_swap}. Then, \textbf{Swap} is applied between the buckets, and thereafter we check whether the fairness constraint value(denoted by delta) has improved using a fast formula. If an improvement is found, we update the ranking and put the reverse of the items in the \texttt{tabu\_list\_swap}. The \texttt{tabu\_list\_swap} is updated continuously up to a maximum capacity \texttt{tabu\_list\_swap\_max}; once this limit is reached, the oldest entry is removed to accommodate new entries.
At any point, if delta is less than or equal to $\epsilon$ 
, we have obtained a fair ranking. We then calculate the distance between the fair ranking and the original ranking and update the distance tracker, dist**.

The process continues for several iterations as determined by \texttt{max\_number\_swap} and seeks to find new improvements in the fairness constraint and concludes when the no improvement flag counter known as global\_number\_swap is reached.
Following this, we apply \textbf{CreatingTies}  to the buckets, and proceed similarly, except that once an improvement is found, we not only put those buckets in the \texttt{tabu\_list\_createties} but also in  \texttt{tabu\_list\_breakties}.
Lastly, we apply the \textbf{BreakingTies} module to buckets that have multiple items, with the only condition that the bucket has not been produced by the \textbf{CreatingTies} module. We start this module with buckets having smaller sizes and move towards larger sizes.
This is because breaking a smaller bucket will result in a small increment in the distance metric.

We extract one item at a time, placing it in a new bucket and the remaining items in another. We also allow for the swap between the newly formed buckets and then check for improvement in the fairness constraint, and accept the combination that yields the lowest delta;
then we add the new buckets in  \texttt{tabu\_list\_breakties} and \texttt{tabu\_list\_createties}.
 
\subsection*{**Post-Processing Steps}
After updating the distance tracker, we apply the following additional steps to the fair ranking. These steps are applied at the individual item level and not at the bucket level.
\begin{itemize}
\item We allow new swaps to occur between the items within a particular group; the swap that reduces the initial distance is kept, and the ranking is updated.
This process continues in this manner until the reduction in distance stops, at which point we update the global distance tracker and apply the second step described below.
\item We then check for any pair of items that were tied in the original unfair ranking, belonged to the same group, but due to the 'BreakingTies' procedure have been assigned consecutive ranks, and whose consecutive ranks are not shared by any item(s) of the other group. We make them tied again in the fair ranking and assign the minimum of the consecutive ranks to those items. This is important as it reduces distance and practically has no impact on the fairness constraint. After this, the fair ranking is converted back into the bucket level to be used by other modules of the algorithm and the distance tracker is updated globally.
\end{itemize}

\subsection*{Simulated Annealing Phase}
This phase runs in tandem with the above-mentioned section. When a move is applied, and an improvement in the fairness constraint is found, we cannot update the ranking every time, because doing so would enlarge the distance between the original ranking and the fair ranking. To prevent this, we use SA (with a fixed cooling schedule) to decide whether to accept or reject a move.

\subsection*{Epsilon Greedy Policy}
After having obtained a fair ranking through the SLS phase, we attempt to perturb the ranking. We have 2 particular strategies to allow for changes in the solution. 
\begin{enumerate}
    \item Complete Reset
    \item Destroy a part of the ranking
\end{enumerate}

Complete Reset:
We change the state of the algorithm to the initial state; the fairness constraint value and its components (g1\_actual, g2\_actual) are reset to the original state as well, and all the tabu lists are cleared.

Destroy a part of the ranking: We only retain (1-\textit{x})\% of the fair ranking, and the remaining \textit{x}\% of the fair ranking is obtained by shuffling the rank values.
However, which items will retain the same rank value as the fair ranking is decided randomly.
The fairness constraint value is calculated for this newly obtained ranking, and (g1\_actual, g2\_actual) are assigned new values based on this ranking. All the tabu lists are cleared as well.
The \textit{x} parameter is reduced using an exponential formula whenever the distance tracker does not improve.
The Epsilon-Greedy policy is used to decide which perturbation strategy to use in a particular iteration. Each strategy is treated as an action.
If the use of the chosen action reduces the distance value, the reward is set to 1, otherwise 0; and correspondingly, the score of the chosen strategy is updated.

The algorithm accepts twelve hyperparameters which configure its behaviour; details are provided in Table~\ref{tab:placeholder}. The overall procedure is formalised in Algorithm~\ref{alg:complex}. 

\begin{table}[H]
    \centering
    \footnotesize
     \begin{tabular}{lll}
     \toprule
      \textbf{Hyperparameter} & \textbf{Ranges} & \textbf{Purpose} \\
       \midrule
        max\_iterations & [2, n] & \makecell[l]{Maximum iterations allowed\\ for the algorithm}\\
        temperature & (0,1] & \makecell[l]{Initial temperature parameter\\ for Simulated Annealing}\\
        k & [1, (n*(n-1))/2] & \makecell[l]{Maximum bound \\constraining the other \\hyperparameters}\\ 
        tabu\_list\_swap\_max &[1, k]& \makecell[l]{Maximum size of tabu list\\ swap}\\
        tabu\_list\_ct\_max &[1, k]& \makecell[l]{Maximum length of tabu \\list creating of ties}\\
        global\_num\_swap & [1,k] & \makecell[l]{Early Stopping threshold \\ for swap module}\\
        global\_num\_ct & [1,k] & \makecell[l]{Early Stopping threshold \\ for CreatingTies module}\\
        seed & [1,9999] & \makecell[l]{Random seed\\ for reproducibility}\\
        num1 & [10,1000] & \makecell[l]{Constant value\\ to increase temperature}\\
        num2 & [1, max\_iterations] & \makecell[l]{Interval step\\ to increase temperature}\\
        x & [0.75, 1] &  \makecell[l]{Perturbation factor}\\
        $\gamma$ & [0.05,0.3] &  \makecell[l]{Exploration rate for \\Epsilon-Greedy policy}\\
        gnum & [0.1,0.35] &  \makecell[l]{Early-stopping percentage to\\ terminate the algorithm}\\
    \bottomrule  
    \end{tabular}
    \caption{Hyperparameters Description}
    \label{tab:placeholder}
\end{table}


\algrenewcommand\algorithmicreturn{\textbf{return}}
\begin{algorithm}
\caption{Tie-Aware Fair Rank Heuristic}
\label{alg:complex}
\begin{algorithmic}[1]
\State \textbf{Input:} Input dataset $D$,  $\epsilon$, hyperparameters (e.g., max iterations, temperature, etc.)
\State \textbf{Output:} Fair Ranking Dataset $R$
\State $set\_seed(seed)$
\If { $calculate\_fairness\_constraint(D) \le \epsilon$ }
    \State $f \gets D[Rank\_Orig]$  \Comment{Extract column Rank\_Orig}
    \State $R \gets \text{ADD\_COLUMN}(D, \text{"Fair\_Rank"}, f)$
    \State \Return $R$

\Else
    \State $iteration \gets 0$
    
    \While {$iteration < max\_iterations$}
        \State $swap\_execution()$
        \State $create\_ties\_execution()$
         \State $break\_ties\_execution()$

        \State $iteration \gets iteration + 1$
        \State $Epsilon\_greedy\_policy\_perturbation(gamma)$
        \State $interval\_step= \max\left(1,\; \left\lfloor \frac{\text{max\_iterations}}{\text{num2}} \right\rfloor\right)$\\
        \If{$iteration \bmod interval\_step$}\\
           \State $temperature*=num1$
        \EndIf
        \If {{no\_improvement\_flag}$\ge$int(gnum*max\_iterations)}\\
           \Comment{{\footnotesize no\_improvement\_flag is to keep track of no improvement in distance}}
           \State \textbf{break}
        \EndIf
    \EndWhile
\State \Return $R$

\EndIf
\end{algorithmic}
\end{algorithm}

\section{\textbf{Experimentation and Findings}}
\label{sec:experiments}
\subsection*{Exact Algorithm Findings}

For this study, we generate multiple synthetic datasets using Algorithm~\ref{alg:cap1}. Each dataset consists of $n$ items where $5 \leq n \leq 113$, an unfair ranking, and protected attribute information. We used two values for tie\_probability, 0.3 and 0.6, to generate the instances. These values were chosen arbitrarily and bear no particular significance. Specifically, a value of 0.3 was used for instances with $\text{size}<50$, while a value of 0.6 was used for instances with $\text{size}\ge50$. 

Further, we tested the exact algorithm on 10 distinct threshold points to adjust for the degree of fairness. Based on our analysis, we have found that the fair ranking produced by EFTA-BILP may introduce new ties, break original ties, completely change item positions, or even result in a ranking with no ties at all. 

\begin{figure}[htbp]
    \centering
    \begin{subfigure}[b]{0.25\textwidth}
        \centering
        \includegraphics[width=\textwidth]{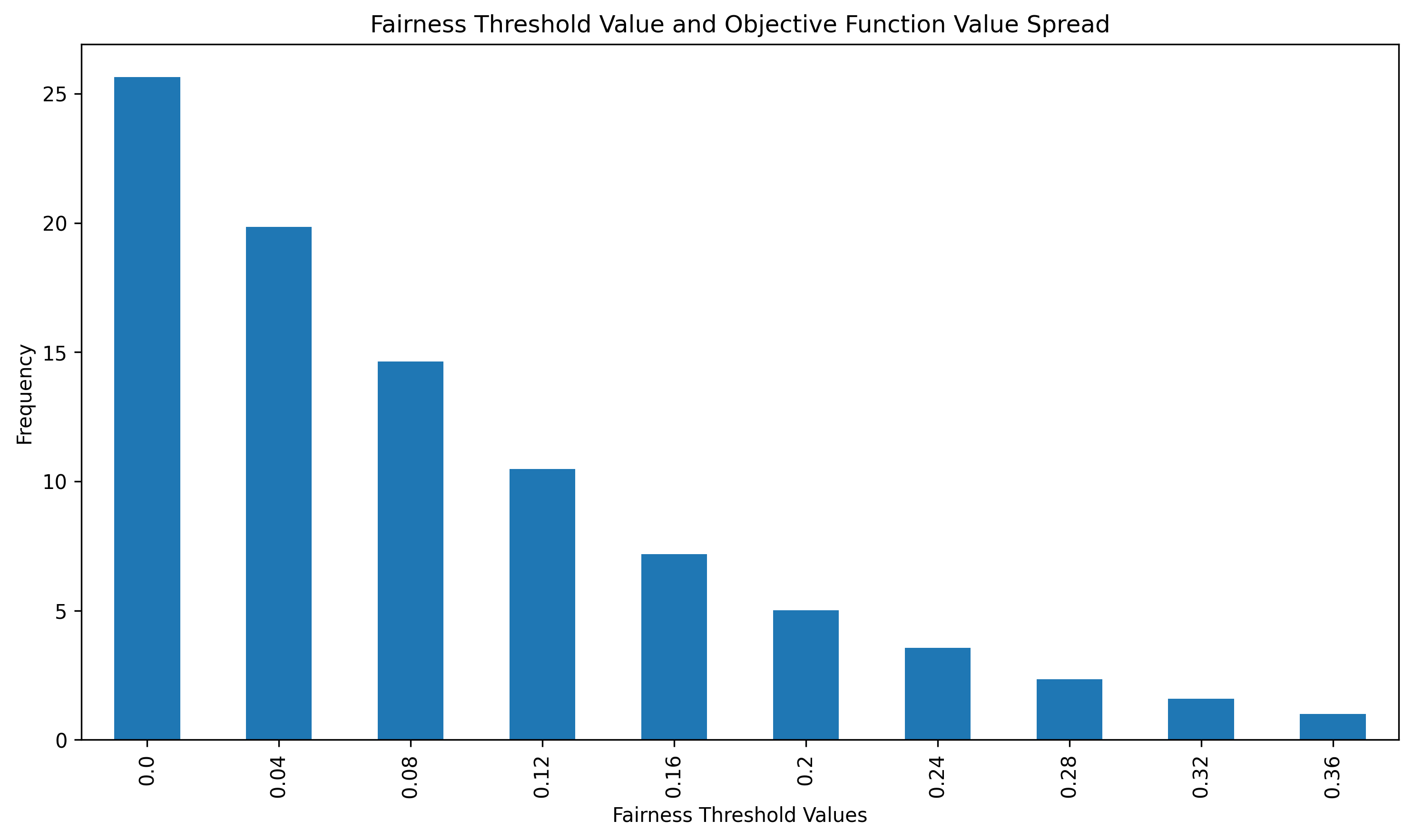}
        \caption{Relationship between distance metric and degree of fairness}
        \label{fig:threshold}
    \end{subfigure}
    \hfill
    \begin{subfigure}[b]{0.25\textwidth}
        \centering
        \includegraphics[width=\textwidth]{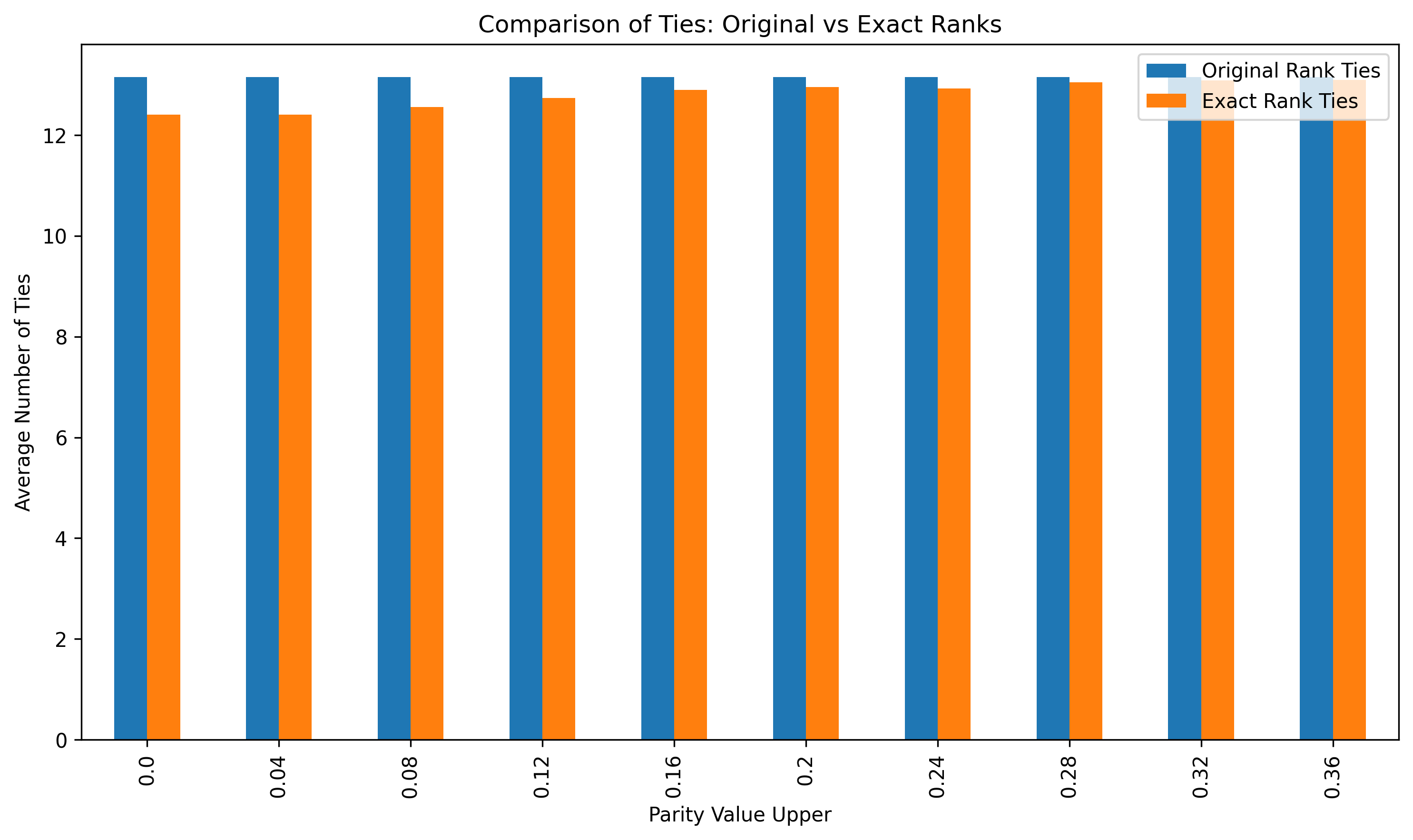}
        \caption{Ties comparison}
        \label{fig:ties}
    \end{subfigure}
\end{figure}

Figure~\ref{fig:threshold} illustrates an intuitive insight: to achieve strict parity (\(\varepsilon = 0\)), the original unfair ranking must be significantly transformed, resulting in maximum distance between the original ranking and the transformed ranking. This distance decreases as the parity constraint is relaxed.

Interestingly, we have never encountered a situation where the algorithm resulted in infeasibility* \footnote{*We have observed that infeasibility can sometimes arise when we restrict the linear program to return a ranking with a number of buckets similar to the original ranking, or to a specific number of buckets} for any threshold value. This is due to the presence of ties, which ensure fairness in every possible scenario, but it can be counterproductive. Adding an excessive number of ties to the ranking pushes it further away from the original ranking. However, the objective function penalises this, so the number of ties added to the ranking remained minimal, as shown in Figure~\ref{fig:ties}.

However, with the current implementation, obtaining solutions for instances with more than 104 items is computationally infeasible within a reasonable time frame.


The algorithm can also handle multiple rankings simultaneously; in that case, it serves as an in-process method. However, we confined our experiments to a single ranking; further experimentation is needed to fully assess the limitations of the in-process approach. 

The algorithm is implemented in Python using the Gurobi Solver \cite{gurobi} under a free academic license and compiled using Cython \cite{behnel2010cython}. For each dataset, different threshold values were evaluated concurrently on a 64-core HPC node.

Further note that the optimal fair ranking returned by EFTA-BILP may not be unique. Even when using the same random seed and identical script, we observed different optimal solutions with the same distance value when running the algorithm in different environments (specifically, on an HPC cluster and a local machine). This discrepancy is likely due to differences in the Gurobi version used in each environment. 
\subsection*{Heuristic Performance}
Algorithm~\ref{alg:complex} follows a slightly different implementation setup from the exact algorithm. For hyperparameter optimisation (HPO), we used Optuna \cite{ozaki2025optunahub}, applying the Tree-Structured Parzen Estimator (TPE) \cite{watanabe2023tree}, a Bayesian optimisation method, within a fixed budget of 300 trials, and tuned the hyperparameters for a given instance and 10 distinct epsilon values and two different seed values. To carry out this efficiently, we created a SLURM job to run all tasks(\#instances*\#epsilon\_values*seed\_values) concurrently.
Note that TPE was applied in a sequential manner to avoid issues with irreproducibility.

For presentation purposes, the data instances were grouped into three size categories, as summarised in Table~\ref{tab:categories}.

\begin{table}[H]
\centering
\caption{Data instance categories by size}
\label{tab:categories}
\begin{tabular}{c c}
\toprule
\textbf{Category} & \textbf{Size} \\
\midrule
Small  & $n < 20$ \\
Medium & $20 \le n < 50$ \\
Large  & $n \ge 50$ \\
\bottomrule
\end{tabular}
\end{table}

Tables~\ref{tab:placeholder2} and~\ref{tab:placeholder3} summarise the principal quantitative indicators used to evaluate the performance of the algorithm. The ``Exact Match'' metric measures the accuracy of the heuristic by calculating the proportion of instances for which it obtains the same distance as the exact algorithm. When an exact match is not found, metrics such as average, minimum, and median reflect the deviations, calculated as (distance - exact distance). 
The results show that the proposed algorithm achieves accuracies of approximately (calculated over two seed values, hence provided the range) \smallAccuracy\% and \mediumAccuracy\% for small and medium instances, respectively. These results indicate strong agreement with the exact algorithm, even under the strict condition of $\epsilon = 0$. However, for large instances, the accuracy is more modest; while the heuristic's result is reasonably close enough to the exact algorithm, its value tends to be slightly higher.

To complement the tabulated results, Figures 2 and 3 visualise the performance of the algorithm for $\epsilon$=0 across different instance sizes. These plots provide a more detailed view of the algorithm's behaviour and its agreement with the exact algorithm for small, medium, and large instances.
\begin{figure}[H]
    \centering
    \includegraphics[width=1\linewidth]{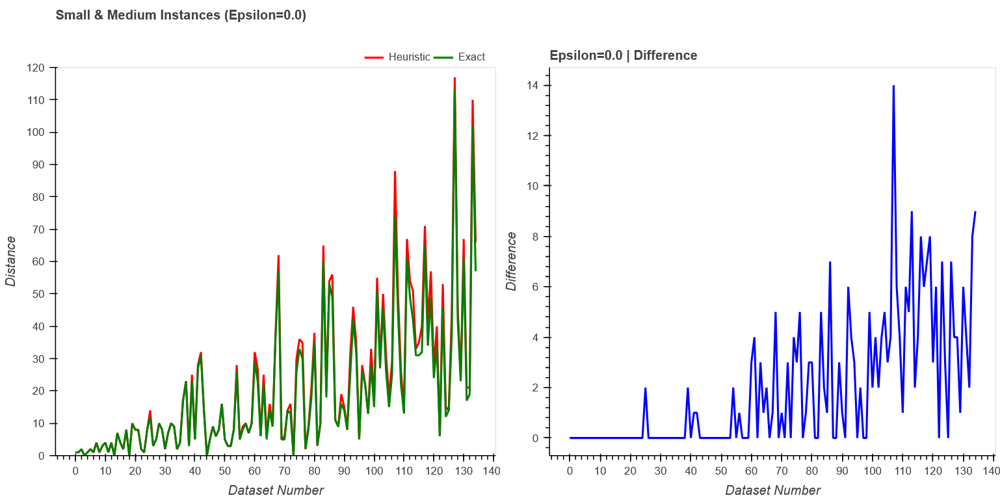}
    \caption{Small and Medium Instances}
    \label{fig:placeholder}
\end{figure}

\begin{figure}[H]
    \centering
    \includegraphics[width=1\linewidth]{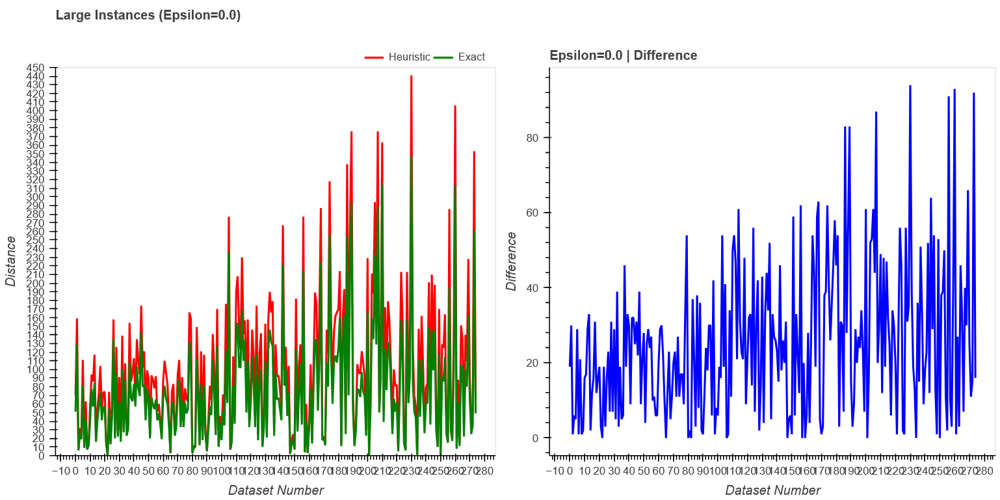}
    \caption{Large Instances}
    \label{fig:placeholder1}
\end{figure}
For $\epsilon>0$, the plots are provided in the Appendix.



\begin{table*}[!htbp]
\noindent
\begin{minipage}{\textwidth}
\caption{Performance metrics for Small/Medium instances}
\centering
\begin{adjustbox}{max width=\textwidth}

\begin{tabular}{lllccccccccccc}
\toprule
Category & Seed Values & Metric & $\epsilon$=0.0 & $\epsilon$=0.04 & $\epsilon$=0.08 & $\epsilon$=0.12 & $\epsilon$=0.16 & $\epsilon$=0.2 & $\epsilon$=0.24 & $\epsilon$=0.28 & $\epsilon$=0.32 & $\epsilon$=0.36 \\
\midrule
Small & 42 & Total & 45.0 & 45.0 & 45.0 & 45.0 & 45.0 & 45.0 & 45.0 & 45.0 & 45.0 & 45.0 \\
Small & 42 & Exact Match & 38.0 & 35.0 & 40.0 & 37.0 & 40.0 & 39.0 & 40.0 & 40.0 & 41.0 & 40.0 \\
Small & 42 & Min variation in distance & 1.0 & 1.0 & 1.0 & 1.0 & 1.0 & 1.0 & 1.0 & 1.0 & 1.0 & 1.0 \\
Small & 42 & Average Variation & 1.57 & 1.2 & 1.4 & 1.75 & 1.2 & 1.83 & 1.4 & 1.4 & 1.5 & 1.6 \\
Small & 42 & Median Variation & 2.0 & 1.0 & 1.0 & 1.0 & 1.0 & 1.5 & 1.0 & 1.0 & 1.0 & 1.0 \\
Small & 42 & Max variation in distance & 2.0 & 2.0 & 3.0 & 5.0 & 2.0 & 4.0 & 2.0 & 2.0 & 3.0 & 3.0 \\
Small & 52 & Total & 45.0 & 45.0 & 45.0 & 45.0 & 45.0 & 45.0 & 45.0 & 45.0 & 45.0 & 45.0 \\
Small & 52 & Exact Match & 39.0 & 37.0 & 37.0 & 34.0 & 41.0 & 38.0 & 42.0 & 40.0 & 42.0 & 40.0 \\
Small & 52 & Min variation in distance & 1.0 & 1.0 & 1.0 & 1.0 & 1.0 & 1.0 & 1.0 & 1.0 & 1.0 & 1.0 \\
Small & 52 & Average Variation & 1.5 & 1.38 & 1.25 & 1.18 & 1.5 & 1.57 & 1.67 & 1.2 & 1.0 & 1.6 \\
Small & 52 & Median Variation & 1.0 & 1.0 & 1.0 & 1.0 & 1.5 & 1.0 & 1.0 & 1.0 & 1.0 & 2.0 \\
Small & 52 & Max variation in distance & 3.0 & 4.0 & 2.0 & 2.0 & 2.0 & 3.0 & 3.0 & 2.0 & 1.0 & 2.0 \\
Medium & 42 & Total & 90.0 & 90.0 & 90.0 & 90.0 & 90.0 & 90.0 & 90.0 & 90.0 & 90.0 & 90.0 \\
Medium & 42 & Exact Match & 26.0 & 30.0 & 43.0 & 58.0 & 66.0 & 71.0 & 75.0 & 81.0 & 85.0 & 85.0 \\
Medium & 42 & Min variation in distance & 1.0 & 1.0 & 1.0 & 1.0 & 1.0 & 1.0 & 1.0 & 2.0 & 3.0 & 1.0 \\
Medium & 42 & Average Variation & 4.72 & 4.95 & 5.23 & 5.38 & 4.17 & 4.63 & 3.67 & 4.22 & 5.2 & 2.6 \\
Medium & 42 & Median Variation & 4.0 & 5.0 & 4.0 & 4.0 & 3.0 & 3.0 & 3.0 & 4.0 & 4.0 & 2.0 \\
Medium & 42 & Max variation in distance & 14.0 & 15.0 & 16.0 & 15.0 & 14.0 & 15.0 & 11.0 & 9.0 & 9.0 & 5.0 \\
Medium & 52 & Total & 90.0 & 90.0 & 90.0 & 90.0 & 90.0 & 90.0 & 90.0 & 90.0 & 90.0 & 90.0 \\
Medium & 52 & Exact Match & 24.0 & 31.0 & 41.0 & 56.0 & 67.0 & 72.0 & 76.0 & 81.0 & 84.0 & 85.0 \\
Medium & 52 & Min variation in distance & 1.0 & 1.0 & 1.0 & 1.0 & 1.0 & 1.0 & 1.0 & 1.0 & 2.0 & 1.0 \\
Medium & 52 & Average Variation & 4.48 & 4.85 & 4.88 & 5.32 & 4.35 & 4.78 & 3.07 & 4.0 & 4.17 & 2.4 \\
Medium & 52 & Median Variation & 4.0 & 4.0 & 4.0 & 4.0 & 3.0 & 3.0 & 2.0 & 3.0 & 3.5 & 2.0 \\
Medium & 52 & Max variation in distance & 15.0 & 14.0 & 15.0 & 21.0 & 14.0 & 15.0 & 7.0 & 10.0 & 9.0 & 4.0 \\
\bottomrule
\end{tabular}
\label{tab:placeholder2}
\end{adjustbox}
\end{minipage}

\vspace{3pt}
\noindent\footnotesize\textit{Note:} The variation metrics are computed considering only the instances where an exact match was not achieved.
\end{table*}

\begin{table*}[!htbp]
\noindent
\begin{minipage}{\textwidth}
\caption{Performance metrics for Large instances}
\centering
\begin{adjustbox}{max width=\textwidth}
\begin{tabular}{lllcccccccccc}
\toprule
Category & Seed Values & Metric & $\epsilon$=0.0 & $\epsilon$=0.04 & $\epsilon$=0.08 & $\epsilon$=0.12 & $\epsilon$=0.16 & $\epsilon$=0.2 & $\epsilon$=0.24 & $\epsilon$=0.28 & $\epsilon$=0.32 & $\epsilon$=0.36 \\
\midrule
Large & 42 & Total & 275.0 & 275.0 & 275.0 & 275.0 & 275.0 & 275.0 & 275.0 & 275.0 & 275.0 & 275.0 \\
Large & 42 & Exact Match & 11.0 & 75.0 & 137.0 & 185.0 & 226.0 & 250.0 & 260.0 & 269.0 & 272.0 & 273.0 \\
Large & 42 & Min variation in distance & 1.0 & 1.0 & 1.0 & 1.0 & 1.0 & 1.0 & 1.0 & 4.0 & 2.0 & 3.0 \\
Large & 42 & Average Variation & 29.05 & 27.29 & 23.71 & 19.6 & 20.63 & 20.08 & 16.33 & 11.33 & 8.33 & 5.0 \\
Large & 42 & Median Variation & 26.0 & 24.0 & 18.0 & 11.0 & 12.0 & 16.0 & 15.0 & 7.5 & 10.0 & 5.0 \\
Large & 42 & Max variation in distance & 94.0 & 104.0 & 99.0 & 108.0 & 88.0 & 58.0 & 37.0 & 28.0 & 13.0 & 7.0 \\
Large & 52 & Total & 275.0 & 275.0 & 275.0 & 275.0 & 275.0 & 275.0 & 275.0 & 275.0 & 275.0 & 275.0 \\
Large & 52 & Exact Match & 11.0 & 76.0 & 138.0 & 183.0 & 226.0 & 249.0 & 260.0 & 268.0 & 272.0 & 273.0 \\
Large & 52 & Min variation in distance & 1.0 & 1.0 & 1.0 & 1.0 & 1.0 & 1.0 & 1.0 & 1.0 & 1.0 & 3.0 \\
Large & 52 & Average Variation & 28.37 & 26.94 & 23.74 & 18.82 & 19.84 & 20.5 & 14.53 & 10.71 & 6.0 & 5.0 \\
Large & 52 & Median Variation & 27.0 & 22.0 & 19.0 & 11.0 & 11.0 & 17.0 & 14.0 & 8.0 & 5.0 & 5.0 \\
Large & 52 & Max variation in distance & 107.0 & 135.0 & 87.0 & 94.0 & 72.0 & 58.0 & 38.0 & 28.0 & 12.0 & 7.0 \\
\bottomrule
\end{tabular}
\label{tab:placeholder3}
\end{adjustbox}
\end{minipage}
\vspace{3pt}
\noindent\footnotesize\textit{Note} The variation metrics are computed considering only the instances where an exact match was not achieved.
\end{table*}

\section{\textbf{Limitations and Future Work}}

Our current approach considers a binary protected attribute, while in practice, a protected attribute can have multiple groups. 
An important direction is extending the current fairness criterion to allow for more than 2 groups and investigating how the exact algorithm performs in this setting, particularly how ties affect its computational complexity.
Further experimentation is required with different fairness measures, especially proportionate fairness.
Additional experimentation is required for an in-process method to see how the exact algorithm handles multiple rankings and to assess its limitations, if any.

\section{\textbf{Conclusion}}
\label{sec:conclusion}
In this study, we proposed a novel post-processing approach to obtain a closest fair ranking when an unfair ranking with ties is provided under a specific fairness notion. As highlighted, considering ties in the rankings significantly increases problem complexity, while their presence helps satisfy the fairness condition at every possible threshold. For $n\le 104$, our exact method can compute an optimal fair ranking efficiently. For larger instances, our heuristic scales well and produces a fair ranking, though not always the one with the lowest distance value.

\section{\textbf{AI Declaration}}
Commercial AI products have been used sparingly; their usage is limited to refining the text of this document and assisting with some parts of the code implementation. All conceptualization, ideation, experimental design, execution, and analysis are solely the work of the authors.

\bibliographystyle{ieeetr}
\renewcommand{\refname}{\textbf{References}}

\bibliography{references}
\appendix
\section{Additional Results}
\begin{figure}
    \centering
    \includegraphics[width=1\linewidth]{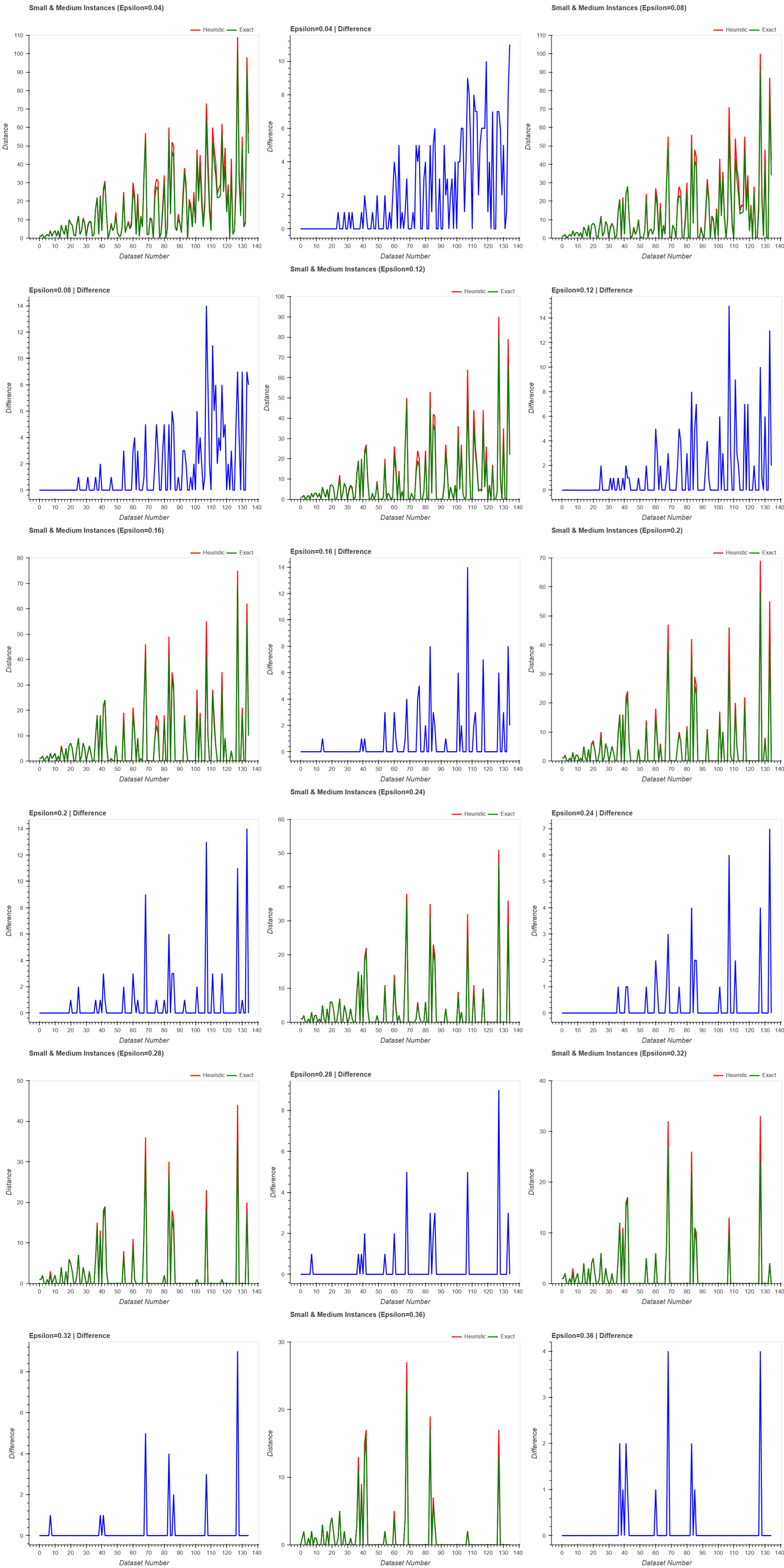}
    \caption{Small/Medium Instances Results for $\epsilon>0$}
    \label{fig:greaterepsilon}
\end{figure}

\begin{figure}
    \centering
    \includegraphics[width=1\linewidth]{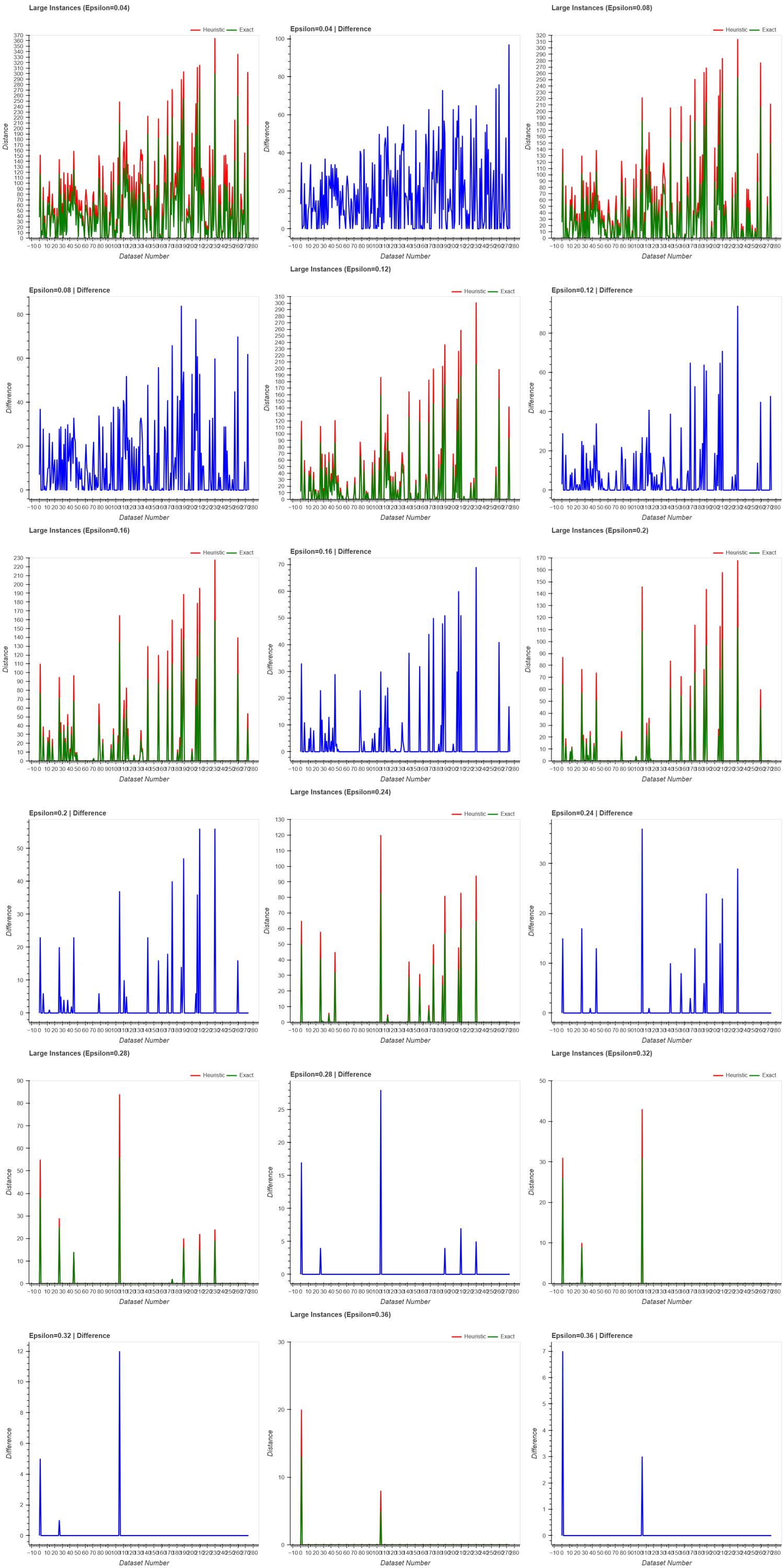}
    \caption{Large Instances Results for $\epsilon>0$}
    \label{fig:greaterepsilon}
\end{figure}

\end{document}